\documentclass[reprint,pra,twocolumn,superscriptaddress]{revtex4-2}
\usepackage[english]{babel}
\usepackage{ucs}
\usepackage[utf8x]{inputenc}
\usepackage{amsmath}
\usepackage{amsfonts}
\usepackage{amssymb}
\usepackage{graphicx}
\usepackage{bbm}
\usepackage{empheq}
\usepackage{color}

\begin{document}
\title{Radio frequency reflectometry in silicon-based quantum dots}
\author{Y.-Y. Liu}
\thanks{These authors contributed equally to this work.}
\affiliation{Department of Physics, Harvard University, Cambridge, Massachusetts 02138, USA}
\author{S. G. J. Philips}
\thanks{These authors contributed equally to this work.}
\affiliation{QuTech and the Kavli Institute of Nanoscience, Delft University of Technology, Delft, 2600, GA, The Netherlands}
\author{L. A. Orona}
\affiliation{Department of Physics, Harvard University, Cambridge, Massachusetts 02138, USA}
\author{N. Samkharadze}
\affiliation{QuTech and Netherlands Organization for Applied Scientific Research (TNO), 2628 CJ Delft, The Netherlands}
\author{T. McJunkin}
\affiliation{University of Wisconsin-Madison, Madison, 53706, Wisconsin, USA}
\author{E. R. MacQuarrie}
\affiliation{University of Wisconsin-Madison, Madison, 53706, Wisconsin, USA}
\author{M. A. Eriksson}
\affiliation{University of Wisconsin-Madison, Madison, 53706, Wisconsin, USA}
\author{L. M. K. Vandersypen}
\email{Correspondence authors: \\
l.m.k.vandersypen@tudelft.nl, yacoby@g.harvard.edu}
\affiliation{QuTech and the Kavli Institute of Nanoscience, Delft University of Technology, Delft, 2600, GA, The Netherlands}
\author{A. Yacoby}
\email{Correspondence authors: \\
l.m.k.vandersypen@tudelft.nl, yacoby@g.harvard.edu}
\affiliation{Department of Physics, Harvard University, Cambridge, Massachusetts 02138, USA}

\begin{abstract}
RF reflectometry offers a fast and sensitive method for charge sensing and spin readout in gated quantum dots. We focus in this work on the implementation of RF readout in accumulation-mode gate-defined quantum dots, where the large parasitic capacitance poses a challenge. We describe and test two methods for mitigating the effect of the parasitic capacitance, one by on-chip modifications and a second by off-chip changes. We demonstrate that these methods enable  high-performance charge readout  in Si/SiGe quantum dots, achieving a fidelity of 99.9\% for a measurement time of 1 $\mu$s.
\end{abstract}

	\maketitle
	
	\section{Introduction} 
	\label{sec:intro}
	 Quantum computing promises significant speedup of computational tasks that are practically impossible to solve on conventional computers \cite{somma2008quantum, carleo2017solving, aspuru2005simulated, shor1999polynomial}.  Of the physical platforms available, spin-based quantum bits (qubits) in semiconductors are particularly promising \cite{Morton11, vandersypen2017interfacing}. 
	 Single-qubit gates with fidelities above 99.9\% \cite{Takeda2018} and two qubit gate fidelities up to 98\% \cite{Huang2019,xue2019benchmarking} have been demonstrated.  Spin qubits in silicon are considered a strong candidate for realizing a large-scale quantum processor due to the small qubit dimensions, localized nature of the control, CMOS compatibility, long coherence times \cite{zwanenburg2013silicon}  and possibility of operating beyond 1 Kelvin \cite{Petit2020, Yang2020}. 
	
	Charge sensing is an important technique for measuring spin qubits as their long-lived spin states can be converted into detectable charge states \cite{elzerman2004single, barthel2009rapid}. 
	To detect a charge state, a sensing dot (SD) is placed in close proximity (d $< \sim$  300 nm) to the qubit. The sensing dot's resistance is strongly dependent on the charge state. 	
	However, measuring this resistance in DC with a amplifier at room temperature requires an integration time on the order of 30 $\mu$s -- 1 ms due to the presence of noise and the RC time constant from the line capacitance and the amplifier input impedance. This slow readout forms a bottleneck when performing spin qubit experiments, since initialization and manipulation can be performed on the nanosecond or microsecond scale \cite{petta2005coherent,koppens2006driven, kim2015high}. 	
	
	Radio Frequency (RF) reflectometry~\cite{Schoelkopf1998} has been successfully applied to depletion-mode GaAs quantum dots and has enabled single shot readout with only several microseconds of integration time~\cite{reilly2007fast}. 
	However, in accumulation-mode devices, the large parasitic capacitance of the accumulation gates to the two dimensional electron gas (2DEG) below provides a low-impedance leakage pathway to ground for the RF signal, complicating RF reflectometry measurements.
	Previous works have addressed this problem by the use of circuit board elements \cite{Volk2019} and careful gate design \cite{Connors2020,noiri2020radio}.

	In this work, we further develop the theoretical model of the leakage pathway introduced by this parasitic capacitance and introduce an on-chip method that effectively removes these parasitic capacitances. We first apply this model in the ``Ohmic-style" implementation, similar to GaAs, where the signal is sent through an ohmic contact.  For this approach, we mitigate the effects of the capacitance by optimizing the onboard elements and sample design.  Later we present the ``split-gate style", where the RF signal is carried by a gate which is capacitively coupled to the 2DEG~\cite{Volk2019}. The leakage pathway to the Ohmic contact is blocked by a resistive channel. 
	
	\section{RF Reflectometry}
	\label{sec:rf_reflectometry}

	\begin{figure*}
		\includegraphics[width = 2\columnwidth]{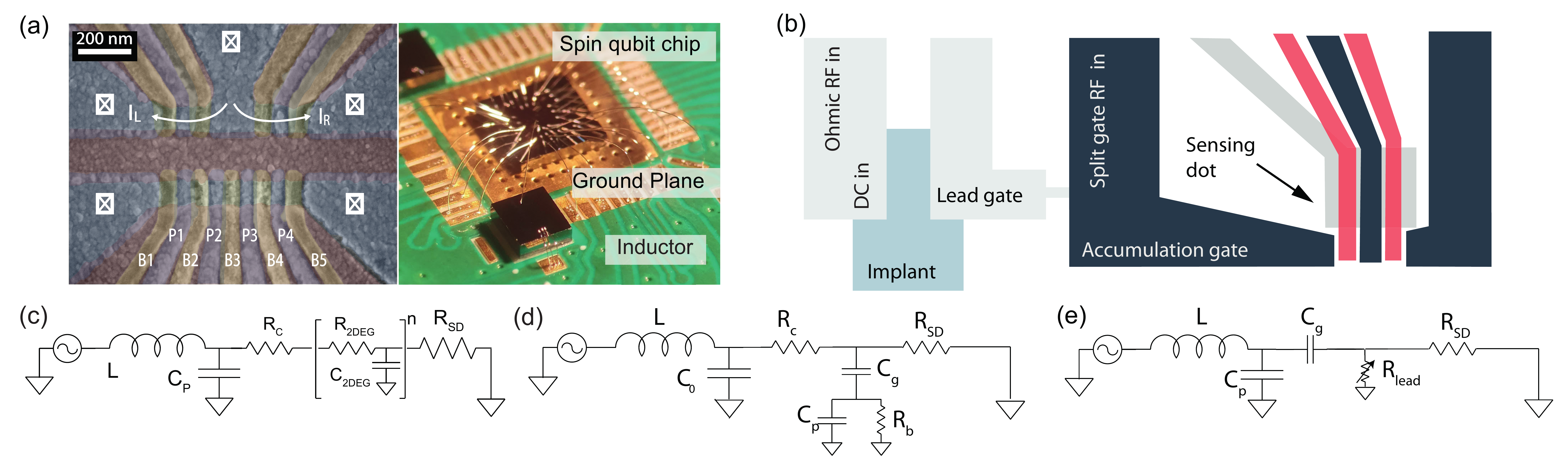}
		\caption{(a) Left: False-colored image of a scanning electron micrograph of a typical Si/SiGe overlapping-gate device. Right: Sample mounted and wire-bonded to a circuit board.  (b) Sketch of the signal path for two approaches. For the ohmic approach the signal is applied to the SD through the ohmic contact and for the split accumulation gate approach the signal is sent through the accumulation gate.  (c) Circuit diagram for the standard method used for GaAs devices.  The bracketed section represents the distributed capacitance and resistance of the 2DEG below the gate.  (d) Circuit diagram for the ohmic method.  The distributed capacitance $C_{2DEG}$ and resistance $R_{2DEG}$ are replaced with lumped elements and absorbed into $C_g$ and $R_c$ for computational simplicity. (e) Circuit diagram for the split accumulation gate approach.}
		\label{fig:overview}
	\end{figure*}
	
	In RF reflectometry a fixed-frequency signal is reflected off an impedance-matching inductive-capacitive (LC) tank circuit that is loaded with the SD with resistance $R_{\rm SD}$, as shown in Figure\ \ref{fig:overview}(c). In the experimental setup $L$ is a lumped element inductor and $C_0$ represents the total capacitance to ground of the circuit board, a lumped element capacitor and the parasitic capacitance of the bond wires and the metal line connecting the bond pad to the implant region close to the quantum dot. The reflection coefficient of this circuit is given by $\Gamma=(Z-Z_0)/(Z+Z_0)$, where $Z$ is the impedance of the loaded tank circuit and $Z_0 = 50 \Omega$ is the source impedance. When the device's parasitic capacitance $C_{\rm 2DEG}$, the contact resistance $R_c$ and the resistance of the reservoir $R_{\rm 2DEG}$ can be ignored, the effective impedance of the loaded tank circuit is $Z = i2\pi fL + 1/(1/(R_{SD} + i2\pi fC_0)$ at an input frequency $f$. 
	
	$\Gamma$ is strongly modulated near $\Gamma = 0$, which occurs at the matching condition $Z=Z_0$ when driven with $f$ equal to the resonance frequency $f_M=1/(2\pi \sqrt{LC_0})$ and with $R_{SD}$ equal to the matching resistance $R_M=L/C_0Z_0$. $R_{SD}$ is very sensitive to the electric potential at the SD and thus to the charges present in the nearby quantum dots. $L$ and $C_0$ are chosen such that the value of $R_M$ matches that where $R_{SD}$ is most sensitive to the qubit dot's charge occupation, typically in the range  of 50 -- 500 k$\Omega$. 

	In Si/SiGe quantum dots, this simple LCR circuit model fails because $R_{\rm 2DEG}$ and $C_{\rm 2DEG}$ are not negligible. Figure\ \ref{fig:overview}(a) shows a typical device (left panel) and circuit board onto which the device is glued (right panel).  A quadruple quantum dot for qubits is formed with the lower set of gates and two sensors are formed with the upper gates. Large accumulation gates control the electron density of the leads from the Ohmic contacts to the quadruple dots and to the SDs. $C_{\rm 2DEG}$ between the 2DEG and the accumulation gates adds up to 0.1--1 pF in total.  The RF signal passes through the lossy 2DEG channel with resistance $R_{\rm 2DEG}$ and is shunted to ground through $C_{\rm 2DEG}$ as shown in Figure\ \ref{fig:overview}(c), drastically reducing the sensitivity of $\Gamma$ to $R_{SD}$.
	
	To overcome this problem, we demonstrate two alternative approaches as shown in Figure\ \ref{fig:overview}(b):
	\begin{itemize}
		\item Ohmic approach -- the RF signal is sent through the Ohmic contacts. The effect of $C_{\rm 2DEG}$ and $R_{\rm 2DEG}$ is mitigated by optimizing the circuit board and sample design (Figure\ \ref{fig:overview}(d)). 
		
		\item Split accumulation gate approach -- the accumulation gate is split into two parts and the RF signal is sent to the SD through the large accumulation gate capacitively (Figure\ \ref{fig:overview}(e)).  $C_{\rm 2DEG}$ becomes the path of the signal rather than a leakage channel. 
	\end{itemize}
	We note that devices with split accumulation gates are used in the experiments demonstrating both methods below. The gate labeled ‘accumulation gate’ serves to accumulate the 2DEG used as the source reservoir for the sensing dot and the ‘lead gate’ accumulates a second 2DEG, connecting the reservoir 2DEG to the ohmic contact.

	\section{Ohmic Approach} 
	
	\label{sub:ohmic_approach_circuit_and_model}
	
	The ohmic approach is shown in Figure\ \ref{fig:overview}(b) and introduces the RF signal to the lead of the SD through the ohmic contact.  The large $C_{\rm 2DEG}$ and $R_{\rm 2DEG}$ prevents applying the simple RLC model to accumulation-mode SiGe devices. For simplicity, the distributed $C_{\rm 2DEG}$ and $R_{\rm 2DEG}$ in Figure\ \ref{fig:overview}(c) can be treated as a single capacitance, $C_g$, and single  resistance, $R_c$ as shown in Figure\ \ref{fig:overview}(d). The gates are connected to ground by two channels: the line resistance $R_{\rm b}$ to  the DC voltage supply which serves as RF ground, and the parasitic capacitance $C_p$ to ground from all the metal on the sample side of $R_b$ (gate, bond wire, bond pad, PCB trace). We will begin by exploring how the tank circuit parameters ($R_b$, $C_p$, $C_0$ and $L$) and the device parameters ($R_c$ and $C_g$) affect the matching conditions ($f_M$ and $R_M$).  This understanding will then be applied to demonstrate several key strategies that allow for Ohmic-style RF reflectometry in Si/SiGe accumulation-mode devices.  The goal is to achieve $R_M$ and $f_M$ values that are experimentally achievable and to ensure the majority of the power is dissipated in $R_{SD}$ resulting in a usable signal-to-noise ratio (SNR). 
	
	\begin{figure}
		\includegraphics[width = \columnwidth]{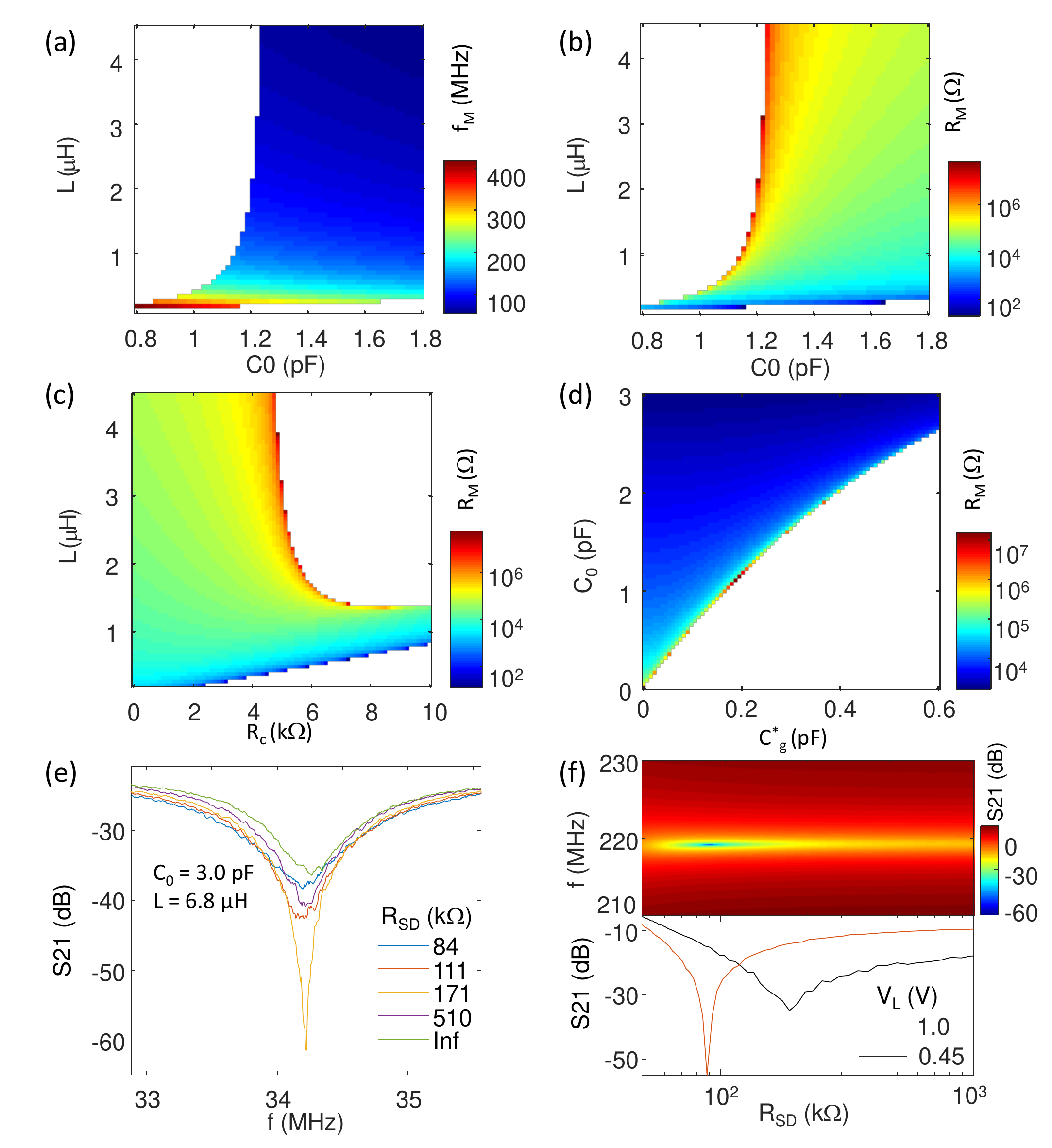}		
		\caption{(a) and (b) Simulations of $f_M$ and $R_M$ as a function of $C_0$ and $L$ with fixed parameters of $C^*_g$ = 0.2 pF and $R_c$ = 3 k$\Omega$. White regions are where no matching can be achieved.  (c) Simulation of $R_M$ as a function of $R_c$ and $L$ with fixed $C^*_g$ = 0.2 pF, and $C_0$ = 1.6 pF. (d) Simulation of $R_M$ as a function of $C_0$ and $C^*_g$ with $L$ = 1 $\mu$H, and $R_c$ = 3  k$\Omega$.  (e) Experimental demonstration of best matching with $f_M$ = 34 MHz and $R_M = 170$ k$\Omega$.  (f) Upper panel: S21 measured as a function of $R_{\rm SD}$ and $f$ for an optimized device and circuit board. Lower panel: S21 at $f_M$ (red and black) as a function of $R_{\rm SD}$ when $V_L$=1 V, 0.45 V respectively.
		}
		\label{fig:Harvard}
	\end{figure}
	
	\textit{Prevent shunting to ground through $C_g$.}\cite{Volk2019} The RF signal in the lead 2DEG has a low impedance path to the accumulation gates through $C_g$.  In order to block this pathway, we have designed the printed circuit board (PCB) to have surface mount resistors to increase $R_b$ between the sample bond pads and the RC filters. $C_p$ is in parallel to $R_b$ and limits the ability to decrease the impact of $C_g$ by just increasing $R_b$.  We place the blocking resistors close to the bond pads to minimize the amount of metal on the sample side and thereby reduce $C_p$. In the end we find a minimum $C_p = 0.2$ pF. The role of $C_g$, $R_b$ and $C_p$ together can be represented by one effective gate capacitor $C^*_g = C_p C_g/(C_p+C_g) = $ 0.2 pF for any $R_b > 100$ k$\Omega$. 
	
	\textit{Solution of Lumped Element Model}. The simple LCR model always has a physically meaningful solution of $f_M$ and $R_M$ for the impedance matching condition.  However, device simulations and experiments demonstrate that large values of $R_c$ and $C^*_g$ can prevent there being a $R_M$ and $f_M$ and therefore the ability to use the tank circuit for charge detection.  In Figure\ \ref{fig:Harvard} we explore the dependence of the matching conditions on $C_0$, L, $C^*_g$ and $R_c$.  Simulations are performed by solving for $R_{\rm SD}$ and $f$ such that the circuit impedance $Z$ matches $Z_0 = 50 \Omega$, giving $R_M$ and $f_M$ respectively.  The constraints that $f_M$ is real and that $R_M$ is real and positive result in there being conditions where no matching can be achieved, which are shown as white regions in Figure\ \ref{fig:Harvard}(a-d). 
	
	\textit{Control matching with $C_0$ and $L$.}  When a sample is fabricated, $C_g$ and $R_c$ are roughly fixed, meaning that the only way to change $R_M$ and $f_M$ is through the tank circuit parameters $L$ and $C_0$.  We present solutions of $f_M$ in Figure\ \ref{fig:Harvard}(a) and $R_M$ in Figure \ \ref{fig:Harvard}(b) as a function of $L$ and $C_0$ with $C^*_g$ = 0.2 pF and $R_c$ = 3 k$\Omega$.    We note that far from the non-matching regions, the behavior is approximately that of the standard LCR model.  Under these conditions, $C_0\gg C_g$ which means that $C_0$ dominates the capacitance of the loaded tank circuit.  When $C_0$ is comparable to or smaller than $C^*_g$, $R_M$ diverges.   
	
	In order to tune $C_0$ and $L$, our PCB has been designed with solder pads for a surface mount inductor, $L$, and a surface mount capacitor to control $C_0$.  
	The board parasitic capacitance also provides a significant contribution ($\sim$ 1 pF) to $C_0$ and sets a lower bound for possible values of $C_0$.  The ground plane near the tank circuit should be minimized to reduce this board parasitic capacitance, ensuring the tunability of the tank circuit by $C_0$ and $L$.  
	Following the prediction of the model, we tested lumped elements with $L$ = 6.8 $\mu$H and $C_0$ = 3.0 pF for a device with estimated $R_c = 4$ k$\Omega$ and $C_g = 0.5$ pF ($C^*_g = 0.2$ pF). The result in Figure\ \ref{fig:Harvard}(e) demonstrates impedance matching behavior with a usable $R_M$. However, it comes at a cost of a very low and unusable $f_M$. 
	Practically, we need $C_0$ to be as low as allowed by $C^*_g$ to guarantee a $f_M$ that is above 100 MHz. For this reason, it is important to reduce $C_g$ and thus $C^*_g$.

	\textit{Balancing $C_g$ and $R_c$ in sample design.} The dependence of the matching conditions is strongly dependent on $R_c$, as shown in Figure\ \ref{fig:Harvard}(c).  At $R_c=0$, the model is reduced to the standard LCR model with an effective $C_0^*=C_0+C_g$. The range of $L$ that can achieve matching is drastically reduced as $R_c$ increases, since more rf power would be dissipated by $R_c$ before $R_{\rm SD}$.  Reducing $R_c$ is therefore key to achieving RF reflectometry.  To capture the impact of $C_g$, we present a simulation of the dependence of $R_M$ in on $C^*_g$ and $C_0$ in Figure\ \ref{fig:Harvard}(d).  We again observe that matching is only achieved when $C_0$ is large enough compared to $C_g$.   
	
	The sample design impacts both $C_g$ and $R_c$, both of which we want to minimize, through the length $l$ and width $w$ of the accumulation gate.  Knowing that $C_g\propto lw$ and $R_c\propto l/w$ reveals that decreasing $l$ is ideal for both parameters while decreasing $w$ to improve $C_g$ comes at the cost of increasing $R_c$ and vice versa.  We have found that $w=$5 $\mu$m is sufficient to achieve consistent accumulation for usable $R_c$ without increasing $C_g$ drastically.  In the future we would place Ohmics as close to the SD as possible to limit $l$ as in  \cite{Connors2020}. The optimized result is demonstrated in Figure\ \ref{fig:Harvard}(f), where we plot the reflected power S21 in the upper panel as a function of $f$ and $R_{SD}$. We apply $V_L = 1$ V on the lead gate to fully turn it on and thus minimize $R_c$. With this we achieved both a usable $R_M \sim 100$ k$\Omega$ and $f_M =220$ MHz.

	\textit{Tuning $R_c$.}  To experimentally confirm the dependence of $R_M$ on $R_c$, we make use of the lead gate.  When $V_L$ is small, the lead gate is partially turned on and thus leads to a larger $R_c$. The lower panel in Figure\ \ref{fig:Harvard}(f) shows S21 at $f_M$ as a function of $R_{\rm SD}$ when $V_L = 1$ V and 0.45 V.	The best matching is achieved with 67 k$\Omega$ for the minimized $R_c$, and 200 k$\Omega$ for a larger $R_c$, which agrees with the simulation in Figure\ \ref{fig:Harvard}(c). This tunability also allows the use of fixed $C_0$ and L for general devices as the matching condition of the device can be tuned in situ.  This tunability, however, is not ideal since the larger $R_c$ gets, more energy is lost before the sensor dot, resulting in a reduced signal to noise ratio.

	\section{Split Gate Approach} 

	\label{sub:lead_gate_approach}
	
	\begin{figure}
		\includegraphics[width=\linewidth]{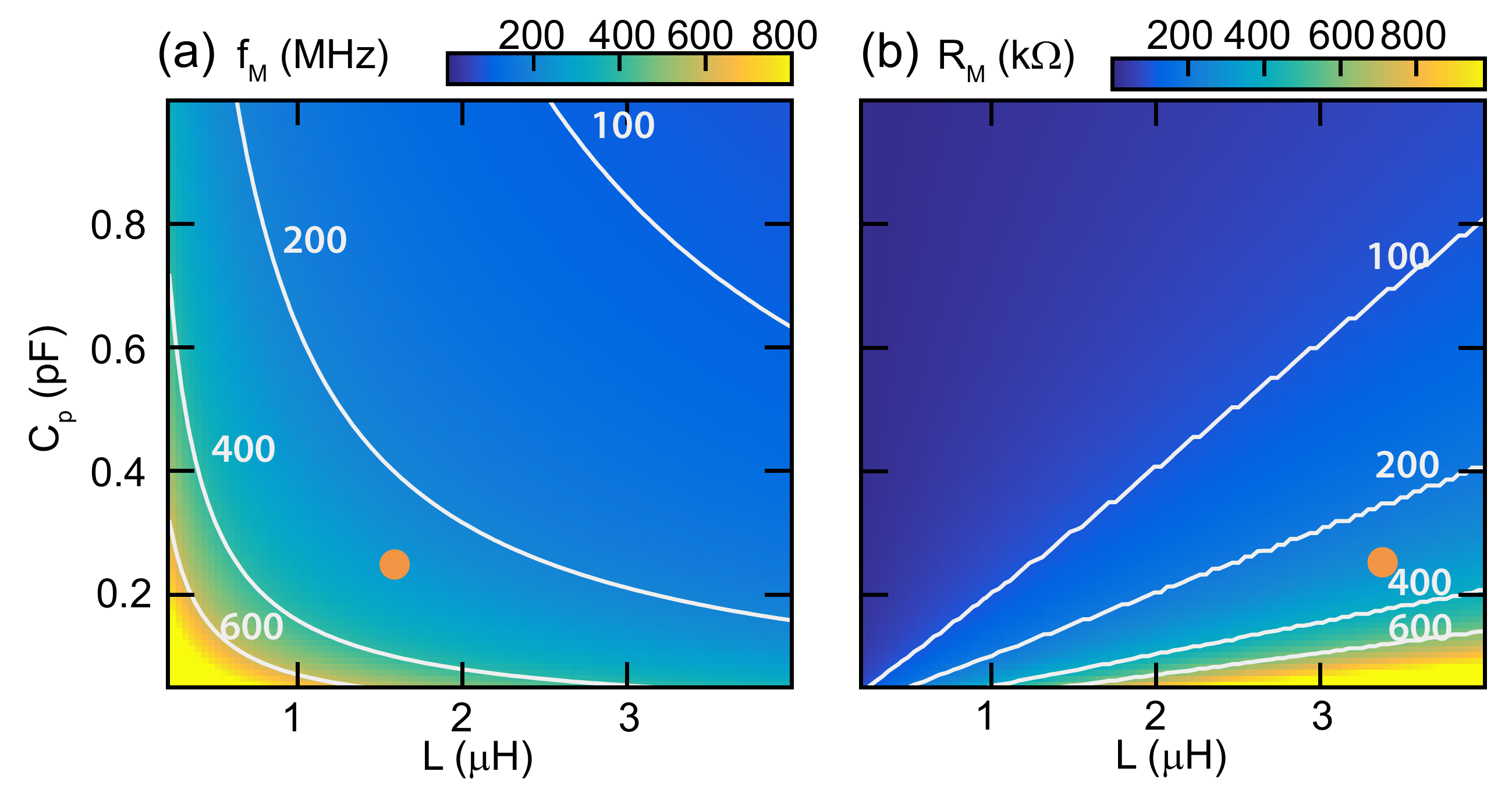}
		\caption{Simulation results for (a) $f_M$ and (b) $R_M$ as a function of $C_p$ and $L$ for the split accumulation gate circuit when $R_{lead}$ = 10 M$\Omega$ and $C_g = 100$ fF. The orange dot indicates the parameters for the device and circuit used in experiment.}
		\label{fig:lead_gate_theory}
	\end{figure}

	In this approach, the RF signal is sent to the sensing dot via the accumulation gate instead of via the Ohmic contact (Figure\ \ref{fig:overview}(b)) \cite{Volk2019}. The capacitance $C_{g}$ between the accumulation gate and 2DEG allows the RF signal to couple in to the 2DEG, as shown in Figure\ \ref{fig:overview}(e). The lead gate is used to generate a high-impedance channel to the Ohmic contact, preventing leakage of the RF signal. 
	
	We aim for similar design specifications for this method as for the Ohmic method: a matching resistance ($R_{M}$) ranging from 200 k$\Omega$ to 600 k$\Omega$ and a resonance frequency ($f_M$) larger than 100 MHz.
	We simulated $f_M$ and $R_{M}$ for different circuit configurations. We estimated $C_g$ to be 100 fF from the sample design and $R_{lead} = 10 \mbox{M}\Omega$. We varied the parasitic capacitance ($C_p$) and the inductance ($L$), as these are parameters controllable by the device design and inductor choice. From the simulation results in Figure\ \ref{fig:lead_gate_theory}, we find a large parameter space that achieves the desired matching condition for practical values of $L$ up to about 5 $\mu$H as long as $C_p < 0.3$ pF. 
	In this case, the circuit reduces to the standard LCR model \cite{taskinen2008radio} given that the reactance of $C_g$, $\chi_g = \frac{1}{2\pi f C_g} << R_{SD}$ and  $R_{lead} >> R_{SD}$. We also simulated the expected measurement bandwidth at the matching condition of this circuit. We only see a weak dependence of the bandwidth on $L$ and $C_p$. The bandwidth of the circuit ranges from 0.5 to 1 MHz.
	
	For the devices used to demonstrate the split accumulation gate approach, we estimated by simulation the total parasitic capacitance to be around $C_p \sim 250$ fF. The parasitic capacitance was kept low using a compact gate layout and high-kinetic-inductance resonators as inductors\cite{Samkharadze2016}. We choose an inductor (resonator) value of L=3.4 $\mu$H, which leads to a resonance frequency of $\sim$180 MHz and a matching resistance of 300 k$\Omega$ for the sensing dot. When operating the device, leakage to the Ohmic contact was cut off by tuning $R_{Lead}$ above 10 M$\Omega$.

	\begin{figure}
		\includegraphics[width=\linewidth]{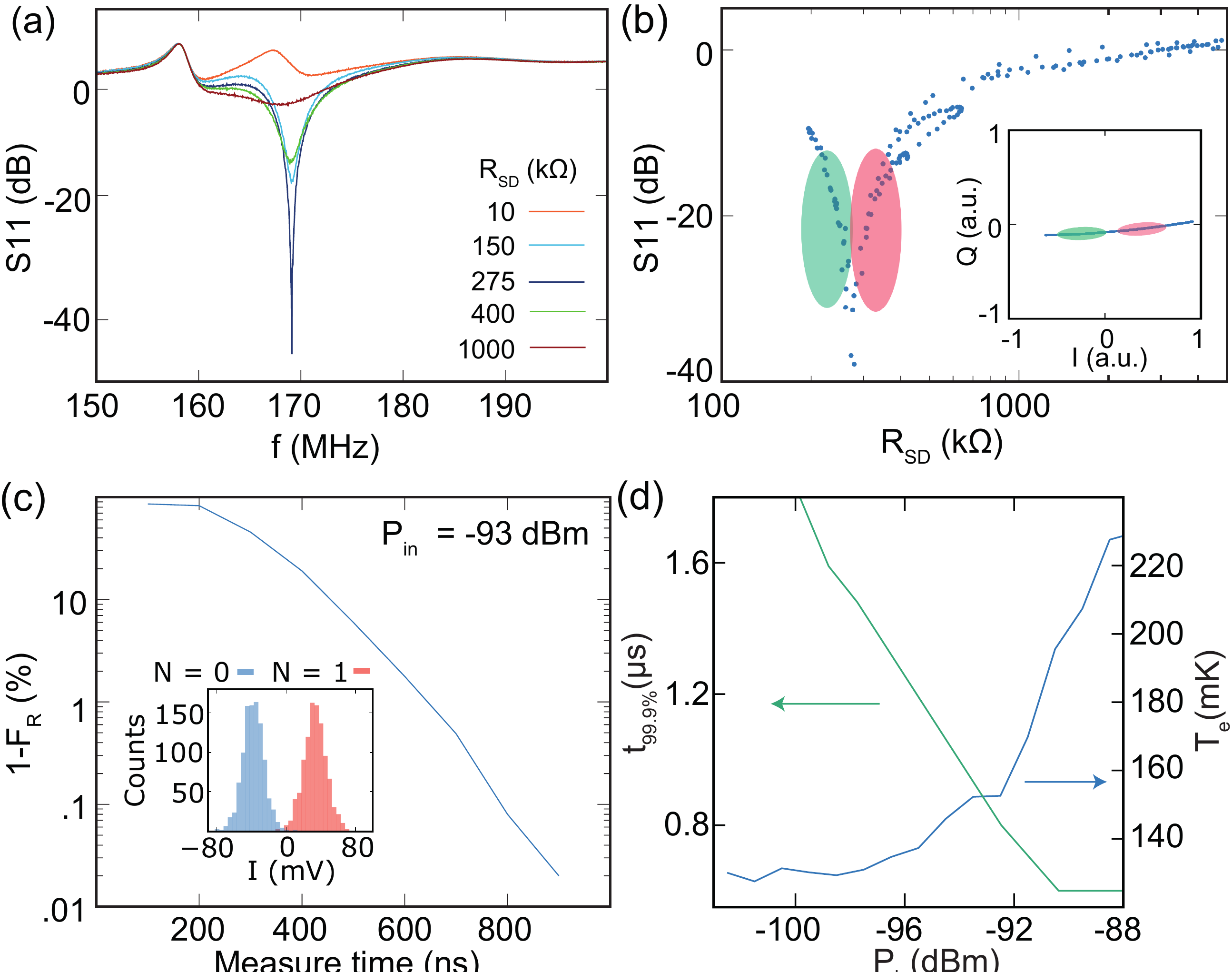}
		\caption{Characteristics obtained with the split accumulation gate approach. (a) Measured reflection coefficient as a function of $f$ for several values of $R_{\rm SD}$. (b) Reflection coefficient $S_{11}$ at $f_{\rm M}$ as a function of $R_{\rm SD}$. $R_{\rm M}=$ 275 k$\Omega$. The sensitive regions are marked in red and green respectively. The inset plots the theoretical response in the IQ plane. (c) Infidelity of charge detection versus measurement time for an interdot transition. The inset: an example histogram for calculating the fidelity. (d) $t_{99.9\%}$ and $T_{\rm e}$ as a function of $P_{\rm in}$.}
		\label{fig:lead_gate_result}
	\end{figure}

	Figure \ref{fig:lead_gate_result}(a) shows the response of the resonator versus frequency for several values of $R_{\rm SD}$, with $f_{\rm M} = 170$ MHz. In Figure\ \ref{fig:lead_gate_result}(b), we find $R_{\rm M} =$ 275 k$\Omega$. The circuit bandwidth can be extracted from the full-width-half-max (FWHM) of the resonance line. For $R_{SD}$ equal to $R_{M}$, the bandwidth is 0.8 MHz which means that we cannot measure signals faster than $\sim$ 600 ns. Two sensitive regions that depend strongly on $R_{SD}$ are visible in Figure\  \ref{fig:lead_gate_result}(b), as indicated by the red and green colored area. The inset shows the expected response of the circuit in the IQ plane around $R_{\rm M}$. The red and green region can by differentiated by a phase $\pi$ in the measured signal. In practice, the coax line between the sample and the measurement circuit adds an unknown phase. In order to  maximize the signal-to-noise ratio (SNR), we record both I and Q and convert the result to a scalar. 
	The resistance $R_{\rm SD}$ was roughly 400 k$\Omega$--1 M$\Omega$, just above $R_{\rm M} =$ 275 k$\Omega$. This means we could improve the SNR by a factor 2 by reducing $C_p$ from 250 fF to 150 fF (smaller gate footprint) or increasing the inductance $L$. This would allow for a value of $R_{\rm M}$ around 600 k$\Omega$.
	
	To characterize the readout performance, we measured the charge readout fidelity ($F_{\rm R}$). This fidelity is defined as the probability to correctly determine whether a quantum dot is occupied with no ($N$=0) or one ($N$=1) electron. To estimate $F_{\rm R}$, we send a train of 10,000 square pulses to the target quantum dot. The dot-reservoir tunnel time is several orders of magnitude shorter than the periods used in the experiment, which means the quantum dot charge state tracks the square pulse. We sample the $IQ$ signal for each half period of the square pulse and plot the distribution for both half periods as shown in the inset of Figure\ \ref{fig:lead_gate_result}(c). The overlap of both signals is the reported infidelity (1-$F_{\rm R}$). 
	For these measurements, we used a digital filter (FIR type) with a passband between 100 kHz and 2.5 MHz. The lower frequency of the passband was determined by the slowest signal we wanted to detect (5 $\mu$s in this case). The upper frequency was taken larger than the bandwidth of the matching circuit to not limit the measurement speed.

	Figure \ref{fig:lead_gate_result}(c) plots the readout infidelity $1-F_{\rm R}$  versus the measurement time when we apply an input signal power ($P_{in}$) of -93 dBm to the readout circuit. We find a minimum measurement time of $t_{99.9\%} =$ 780 ns in order to achieve $F_{\rm R} >  99.9\%$. 
	We see that $t_{99.9\%}$ strongly depends on the input power of the RF-readout circuit (Figure\ \ref{fig:lead_gate_result}(d)). The SNR is improved by larger $P_{in}$ until the bandwidth limit of the circuit is reached (0.8 MHz).
	On the other hand, larger $P_{\rm in}$ also affects the effective electron temperature of the quantum dots. To characterize the trade off, we measured $T_e$ by measuring the polarization line of a interdot transition \cite{van2018automated} and plot the result as a function of $P_{in}$ in Figure\  \ref{fig:lead_gate_result}(d). We note that $T_e$ starts to increase dramatically once $P_{\rm in} >$ -93 dBm. We recommend to only supply power to the RF readout circuit when readout is being done, to prevent the readout from affecting qubit operations.
	
	\section{Conclusion}
	In this work we demonstrated two methods that can be used to achieve a reasonable matching condition for RF reflectometry measurements in accumulation mode devices. For the Ohmic method, we demonstrate that circuit board design can be used to minimize the parasitic capacitance. A careful sample design is necessary in order to obtain both a workable frequency and matching resistance.
	For the split accumulation gate method, the RF source is directed to the accumulation gate of the sensing dot, and the addition of the lead gate allows to efficiently cut off the leakage path to the Ohmic contact. The charge state of a qubit dot can be read out within $1\mu s$ with a $>$99.9\% fidelity, which matches state-of-the-art readout performance. Compared to existing methods demonstrated in recent works~\cite{Connors2020, noiri2020radio}, the split accumulation gate method is especially useful when it is difficult to achieve very low $C_g$ and/or to keep $R_{2DEG}$ sufficiently low. 
    
    \section{Acknowledgements}
    {We acknowledge Lisa Edge from HRL Laboratories for the growth and distribution of the Si/SiGe heterostuctures that were used in this experiment.}
    We acknowledge useful discussions with R. Schouten and the members of our research groups.
    Research was sponsored by the the Army Research Office (ARO), and was accomplished under Grant Number W911NF-17-1-0274. The views and conclusions contained in this document are those of the authors and should not be interpreted as representing the official policies, either expressed or implied, of the Army Research Office (ARO),or the U.S. Government. The U.S. Government is authorized to reproduce and distribute reprints for Government purposes notwithstanding any copyright notation herein.
    We acknowledge financial support from the Marie Skłodowska-Curie actions—Nanoscale solid-state spin systems in emerging quantum technologies—Spin-NANO, grant agreement number 676108.
    The Wisconsin Centers for Nanoscale Technology acknowledges financial support from NSF (DMR–1625348) in the acquisition of the Elionix ELS G-100 electron beam lithography instrument used for part of this work.
    \bibliographystyle{apsrev_lyy}
    \bibliography{SiGe_ref_v1}

\begin{thebibliography}{25}
\expandafter\ifx\csname natexlab\endcsname\relax\def\natexlab#1{#1}\fi
\expandafter\ifx\csname bibnamefont\endcsname\relax
  \def\bibnamefont#1{#1}\fi
\expandafter\ifx\csname bibfnamefont\endcsname\relax
  \def\bibfnamefont#1{#1}\fi
\expandafter\ifx\csname citenamefont\endcsname\relax
  \def\citenamefont#1{#1}\fi
\expandafter\ifx\csname url\endcsname\relax
  \def\url#1{\texttt{#1}}\fi
\expandafter\ifx\csname urlprefix\endcsname\relax\def\urlprefix{URL }\fi
\providecommand{\bibinfo}[2]{#2}
\providecommand{\eprint}[2][]{\url{#2}}

\bibitem[{\citenamefont{Somma et~al.}(2008)\citenamefont{Somma, Boixo, Barnum,
  and Knill}}]{somma2008quantum}
\bibinfo{author}{\bibfnamefont{R.~D.} \bibnamefont{Somma}},
  \bibinfo{author}{\bibfnamefont{S.}~\bibnamefont{Boixo}},
  \bibinfo{author}{\bibfnamefont{H.}~\bibnamefont{Barnum}}, \bibnamefont{and}
  \bibinfo{author}{\bibfnamefont{E.}~\bibnamefont{Knill}}, Quantum simulations
  of classical annealing processes, \bibinfo{journal}{Physical review letters}
  \textbf{\bibinfo{volume}{101}}, \bibinfo{pages}{130504}
  (\bibinfo{year}{2008}).

\bibitem[{\citenamefont{Carleo and Troyer}(2017)}]{carleo2017solving}
\bibinfo{author}{\bibfnamefont{G.}~\bibnamefont{Carleo}} \bibnamefont{and}
  \bibinfo{author}{\bibfnamefont{M.}~\bibnamefont{Troyer}}, Solving the quantum
  many-body problem with artificial neural networks, \bibinfo{journal}{Science}
  \textbf{\bibinfo{volume}{355}}, \bibinfo{pages}{602} (\bibinfo{year}{2017}).

\bibitem[{\citenamefont{Aspuru-Guzik et~al.}(2005)\citenamefont{Aspuru-Guzik,
  Dutoi, Love, and Head-Gordon}}]{aspuru2005simulated}
\bibinfo{author}{\bibfnamefont{A.}~\bibnamefont{Aspuru-Guzik}},
  \bibinfo{author}{\bibfnamefont{A.~D.} \bibnamefont{Dutoi}},
  \bibinfo{author}{\bibfnamefont{P.~J.} \bibnamefont{Love}}, \bibnamefont{and}
  \bibinfo{author}{\bibfnamefont{M.}~\bibnamefont{Head-Gordon}}, Simulated
  quantum computation of molecular energies, \bibinfo{journal}{Science}
  \textbf{\bibinfo{volume}{309}}, \bibinfo{pages}{1704} (\bibinfo{year}{2005}).

\bibitem[{\citenamefont{Shor}(1999)}]{shor1999polynomial}
\bibinfo{author}{\bibfnamefont{P.~W.} \bibnamefont{Shor}}, Polynomial-time
  algorithms for prime factorization and discrete logarithms on a quantum
  computer, \bibinfo{journal}{SIAM review} \textbf{\bibinfo{volume}{41}},
  \bibinfo{pages}{303} (\bibinfo{year}{1999}).

\bibitem[{\citenamefont{Morton and Lovett}(2011)}]{Morton11}
\bibinfo{author}{\bibfnamefont{J.~J.} \bibnamefont{Morton}} \bibnamefont{and}
  \bibinfo{author}{\bibfnamefont{B.~W.} \bibnamefont{Lovett}}, {Hybrid
  Solid-State Qubits: The Powerful Role of Electron Spins},
  \bibinfo{journal}{Annual Review of Condensed Matter Physics}
  \textbf{\bibinfo{volume}{2}}, \bibinfo{pages}{189} (\bibinfo{year}{2011}).

\bibitem[{\citenamefont{Vandersypen et~al.}(2017)\citenamefont{Vandersypen,
  Bluhm, Clarke, Dzurak, Ishihara, Morello, Reilly, Schreiber, and
  Veldhorst}}]{vandersypen2017interfacing}
\bibinfo{author}{\bibfnamefont{L.}~\bibnamefont{Vandersypen}},
  \bibinfo{author}{\bibfnamefont{H.}~\bibnamefont{Bluhm}},
  \bibinfo{author}{\bibfnamefont{J.}~\bibnamefont{Clarke}},
  \bibinfo{author}{\bibfnamefont{A.}~\bibnamefont{Dzurak}},
  \bibinfo{author}{\bibfnamefont{R.}~\bibnamefont{Ishihara}},
  \bibinfo{author}{\bibfnamefont{A.}~\bibnamefont{Morello}},
  \bibinfo{author}{\bibfnamefont{D.}~\bibnamefont{Reilly}},
  \bibinfo{author}{\bibfnamefont{L.}~\bibnamefont{Schreiber}},
  \bibnamefont{and}
  \bibinfo{author}{\bibfnamefont{M.}~\bibnamefont{Veldhorst}}, Interfacing spin
  qubits in quantum dots and donors, hot, dense, and coherent,
  \bibinfo{journal}{npj Quantum Information} \textbf{\bibinfo{volume}{3}},
  \bibinfo{pages}{1} (\bibinfo{year}{2017}).

\bibitem[{\citenamefont{Takeda et~al.}(2018)\citenamefont{Takeda, Yoneda,
  Otsuka, Nakajima, Delbecq, Allison, Hoshi, Usami, Itoh, Oda
  et~al.}}]{Takeda2018}
\bibinfo{author}{\bibfnamefont{K.}~\bibnamefont{Takeda}},
  \bibinfo{author}{\bibfnamefont{J.}~\bibnamefont{Yoneda}},
  \bibinfo{author}{\bibfnamefont{T.}~\bibnamefont{Otsuka}},
  \bibinfo{author}{\bibfnamefont{T.}~\bibnamefont{Nakajima}},
  \bibinfo{author}{\bibfnamefont{M.~R.} \bibnamefont{Delbecq}},
  \bibinfo{author}{\bibfnamefont{G.}~\bibnamefont{Allison}},
  \bibinfo{author}{\bibfnamefont{Y.}~\bibnamefont{Hoshi}},
  \bibinfo{author}{\bibfnamefont{N.}~\bibnamefont{Usami}},
  \bibinfo{author}{\bibfnamefont{K.~M.} \bibnamefont{Itoh}},
  \bibinfo{author}{\bibfnamefont{S.}~\bibnamefont{Oda}}, \bibnamefont{et~al.},
  Optimized electrical control of a Si/SiGe spin qubit in the presence of an
  induced frequency shift, \bibinfo{journal}{npj Quantum Information}
  \textbf{\bibinfo{volume}{4}}, \bibinfo{pages}{54} (\bibinfo{year}{2018}).

\bibitem[{\citenamefont{Huang et~al.}(2019)\citenamefont{Huang, Yang, Chan,
  Tanttu, Hensen, Leon, Fogarty, Hwang, Hudson, Itoh et~al.}}]{Huang2019}
\bibinfo{author}{\bibfnamefont{W.}~\bibnamefont{Huang}},
  \bibinfo{author}{\bibfnamefont{C.~H.} \bibnamefont{Yang}},
  \bibinfo{author}{\bibfnamefont{K.~W.} \bibnamefont{Chan}},
  \bibinfo{author}{\bibfnamefont{T.}~\bibnamefont{Tanttu}},
  \bibinfo{author}{\bibfnamefont{B.}~\bibnamefont{Hensen}},
  \bibinfo{author}{\bibfnamefont{R.~C.~C.} \bibnamefont{Leon}},
  \bibinfo{author}{\bibfnamefont{M.~A.} \bibnamefont{Fogarty}},
  \bibinfo{author}{\bibfnamefont{J.~C.~C.} \bibnamefont{Hwang}},
  \bibinfo{author}{\bibfnamefont{F.~E.} \bibnamefont{Hudson}},
  \bibinfo{author}{\bibfnamefont{K.~M.} \bibnamefont{Itoh}},
  \bibnamefont{et~al.}, Fidelity benchmarks for two-qubit gates in silicon,
  \bibinfo{journal}{Nature} \textbf{\bibinfo{volume}{569}},
  \bibinfo{pages}{532} (\bibinfo{year}{2019}).

\bibitem[{\citenamefont{Xue et~al.}(2019)\citenamefont{Xue, Watson, Helsen,
  Ward, Savage, Lagally, Coppersmith, Eriksson, Wehner, and
  Vandersypen}}]{xue2019benchmarking}
\bibinfo{author}{\bibfnamefont{X.}~\bibnamefont{Xue}},
  \bibinfo{author}{\bibfnamefont{T.}~\bibnamefont{Watson}},
  \bibinfo{author}{\bibfnamefont{J.}~\bibnamefont{Helsen}},
  \bibinfo{author}{\bibfnamefont{D.~R.} \bibnamefont{Ward}},
  \bibinfo{author}{\bibfnamefont{D.~E.} \bibnamefont{Savage}},
  \bibinfo{author}{\bibfnamefont{M.~G.} \bibnamefont{Lagally}},
  \bibinfo{author}{\bibfnamefont{S.~N.} \bibnamefont{Coppersmith}},
  \bibinfo{author}{\bibfnamefont{M.}~\bibnamefont{Eriksson}},
  \bibinfo{author}{\bibfnamefont{S.}~\bibnamefont{Wehner}}, \bibnamefont{and}
  \bibinfo{author}{\bibfnamefont{L.}~\bibnamefont{Vandersypen}}, Benchmarking
  gate fidelities in a Si/SiGe two-qubit device, \bibinfo{journal}{Physical
  Review X} \textbf{\bibinfo{volume}{9}}, \bibinfo{pages}{021011}
  (\bibinfo{year}{2019}).

\bibitem[{\citenamefont{Zwanenburg et~al.}(2013)\citenamefont{Zwanenburg,
  Dzurak, Morello, Simmons, Hollenberg, Klimeck, Rogge, Coppersmith, and
  Eriksson}}]{zwanenburg2013silicon}
\bibinfo{author}{\bibfnamefont{F.~A.} \bibnamefont{Zwanenburg}},
  \bibinfo{author}{\bibfnamefont{A.~S.} \bibnamefont{Dzurak}},
  \bibinfo{author}{\bibfnamefont{A.}~\bibnamefont{Morello}},
  \bibinfo{author}{\bibfnamefont{M.~Y.} \bibnamefont{Simmons}},
  \bibinfo{author}{\bibfnamefont{L.~C.} \bibnamefont{Hollenberg}},
  \bibinfo{author}{\bibfnamefont{G.}~\bibnamefont{Klimeck}},
  \bibinfo{author}{\bibfnamefont{S.}~\bibnamefont{Rogge}},
  \bibinfo{author}{\bibfnamefont{S.~N.} \bibnamefont{Coppersmith}},
  \bibnamefont{and} \bibinfo{author}{\bibfnamefont{M.~A.}
  \bibnamefont{Eriksson}}, Silicon quantum electronics,
  \bibinfo{journal}{Reviews of modern physics} \textbf{\bibinfo{volume}{85}},
  \bibinfo{pages}{961} (\bibinfo{year}{2013}).

\bibitem[{\citenamefont{Petit et~al.}(2020)\citenamefont{Petit, Eenink, Russ,
  Lawrie, Hendrickx, Philips, Clarke, Vandersypen, and Veldhorst}}]{Petit2020}
\bibinfo{author}{\bibfnamefont{L.}~\bibnamefont{Petit}},
  \bibinfo{author}{\bibfnamefont{H.~G.~J.} \bibnamefont{Eenink}},
  \bibinfo{author}{\bibfnamefont{M.}~\bibnamefont{Russ}},
  \bibinfo{author}{\bibfnamefont{W.~I.~L.} \bibnamefont{Lawrie}},
  \bibinfo{author}{\bibfnamefont{N.~W.} \bibnamefont{Hendrickx}},
  \bibinfo{author}{\bibfnamefont{S.~G.~J.} \bibnamefont{Philips}},
  \bibinfo{author}{\bibfnamefont{J.~S.} \bibnamefont{Clarke}},
  \bibinfo{author}{\bibfnamefont{L.~M.~K.} \bibnamefont{Vandersypen}},
  \bibnamefont{and}
  \bibinfo{author}{\bibfnamefont{M.}~\bibnamefont{Veldhorst}}, Universal
  quantum logic in hot silicon qubits, \bibinfo{journal}{Nature}
  \textbf{\bibinfo{volume}{580}}, \bibinfo{pages}{355} (\bibinfo{year}{2020}).

\bibitem[{\citenamefont{Yang et~al.}(2020)\citenamefont{Yang, Leon, Hwang,
  Saraiva, Tanttu, Huang, Camirand~Lemyre, Chan, Tan, Hudson
  et~al.}}]{Yang2020}
\bibinfo{author}{\bibfnamefont{C.~H.} \bibnamefont{Yang}},
  \bibinfo{author}{\bibfnamefont{R.~C.~C.} \bibnamefont{Leon}},
  \bibinfo{author}{\bibfnamefont{J.~C.~C.} \bibnamefont{Hwang}},
  \bibinfo{author}{\bibfnamefont{A.}~\bibnamefont{Saraiva}},
  \bibinfo{author}{\bibfnamefont{T.}~\bibnamefont{Tanttu}},
  \bibinfo{author}{\bibfnamefont{W.}~\bibnamefont{Huang}},
  \bibinfo{author}{\bibfnamefont{J.}~\bibnamefont{Camirand~Lemyre}},
  \bibinfo{author}{\bibfnamefont{K.~W.} \bibnamefont{Chan}},
  \bibinfo{author}{\bibfnamefont{K.~Y.} \bibnamefont{Tan}},
  \bibinfo{author}{\bibfnamefont{F.~E.} \bibnamefont{Hudson}},
  \bibnamefont{et~al.}, Operation of a silicon quantum processor unit cell
  above one kelvin, \bibinfo{journal}{Nature} \textbf{\bibinfo{volume}{580}},
  \bibinfo{pages}{350} (\bibinfo{year}{2020}).

\bibitem[{\citenamefont{Elzerman et~al.}(2004)\citenamefont{Elzerman, Hanson,
  Van~Beveren, Witkamp, Vandersypen, and Kouwenhoven}}]{elzerman2004single}
\bibinfo{author}{\bibfnamefont{J.}~\bibnamefont{Elzerman}},
  \bibinfo{author}{\bibfnamefont{R.}~\bibnamefont{Hanson}},
  \bibinfo{author}{\bibfnamefont{L.~W.} \bibnamefont{Van~Beveren}},
  \bibinfo{author}{\bibfnamefont{B.}~\bibnamefont{Witkamp}},
  \bibinfo{author}{\bibfnamefont{L.}~\bibnamefont{Vandersypen}},
  \bibnamefont{and} \bibinfo{author}{\bibfnamefont{L.~P.}
  \bibnamefont{Kouwenhoven}}, Single-shot read-out of an individual electron
  spin in a quantum dot, \bibinfo{journal}{nature}
  \textbf{\bibinfo{volume}{430}}, \bibinfo{pages}{431} (\bibinfo{year}{2004}).

\bibitem[{\citenamefont{Barthel et~al.}(2009)\citenamefont{Barthel, Reilly,
  Marcus, Hanson, and Gossard}}]{barthel2009rapid}
\bibinfo{author}{\bibfnamefont{C.}~\bibnamefont{Barthel}},
  \bibinfo{author}{\bibfnamefont{D.}~\bibnamefont{Reilly}},
  \bibinfo{author}{\bibfnamefont{C.~M.} \bibnamefont{Marcus}},
  \bibinfo{author}{\bibfnamefont{M.}~\bibnamefont{Hanson}}, \bibnamefont{and}
  \bibinfo{author}{\bibfnamefont{A.}~\bibnamefont{Gossard}}, Rapid single-shot
  measurement of a singlet-triplet qubit, \bibinfo{journal}{Physical Review
  Letters} \textbf{\bibinfo{volume}{103}}, \bibinfo{pages}{160503}
  (\bibinfo{year}{2009}).

\bibitem[{\citenamefont{Petta et~al.}(2005)\citenamefont{Petta, Johnson,
  Taylor, Laird, Yacoby, Lukin, Marcus, Hanson, and
  Gossard}}]{petta2005coherent}
\bibinfo{author}{\bibfnamefont{J.~R.} \bibnamefont{Petta}},
  \bibinfo{author}{\bibfnamefont{A.~C.} \bibnamefont{Johnson}},
  \bibinfo{author}{\bibfnamefont{J.~M.} \bibnamefont{Taylor}},
  \bibinfo{author}{\bibfnamefont{E.~A.} \bibnamefont{Laird}},
  \bibinfo{author}{\bibfnamefont{A.}~\bibnamefont{Yacoby}},
  \bibinfo{author}{\bibfnamefont{M.~D.} \bibnamefont{Lukin}},
  \bibinfo{author}{\bibfnamefont{C.~M.} \bibnamefont{Marcus}},
  \bibinfo{author}{\bibfnamefont{M.~P.} \bibnamefont{Hanson}},
  \bibnamefont{and} \bibinfo{author}{\bibfnamefont{A.~C.}
  \bibnamefont{Gossard}}, Coherent manipulation of coupled electron spins in
  semiconductor quantum dots, \bibinfo{journal}{Science}
  \textbf{\bibinfo{volume}{309}}, \bibinfo{pages}{2180} (\bibinfo{year}{2005}).

\bibitem[{\citenamefont{Koppens et~al.}(2006)\citenamefont{Koppens, Buizert,
  Tielrooij, Vink, Nowack, Meunier, Kouwenhoven, and
  Vandersypen}}]{koppens2006driven}
\bibinfo{author}{\bibfnamefont{F.~H.} \bibnamefont{Koppens}},
  \bibinfo{author}{\bibfnamefont{C.}~\bibnamefont{Buizert}},
  \bibinfo{author}{\bibfnamefont{K.-J.} \bibnamefont{Tielrooij}},
  \bibinfo{author}{\bibfnamefont{I.~T.} \bibnamefont{Vink}},
  \bibinfo{author}{\bibfnamefont{K.~C.} \bibnamefont{Nowack}},
  \bibinfo{author}{\bibfnamefont{T.}~\bibnamefont{Meunier}},
  \bibinfo{author}{\bibfnamefont{L.}~\bibnamefont{Kouwenhoven}},
  \bibnamefont{and}
  \bibinfo{author}{\bibfnamefont{L.}~\bibnamefont{Vandersypen}}, Driven
  coherent oscillations of a single electron spin in a quantum dot,
  \bibinfo{journal}{Nature} \textbf{\bibinfo{volume}{442}},
  \bibinfo{pages}{766} (\bibinfo{year}{2006}).

\bibitem[{\citenamefont{Kim et~al.}(2015)\citenamefont{Kim, Ward, Simmons,
  Savage, Lagally, Friesen, Coppersmith, and Eriksson}}]{kim2015high}
\bibinfo{author}{\bibfnamefont{D.}~\bibnamefont{Kim}},
  \bibinfo{author}{\bibfnamefont{D.~R.} \bibnamefont{Ward}},
  \bibinfo{author}{\bibfnamefont{C.~B.} \bibnamefont{Simmons}},
  \bibinfo{author}{\bibfnamefont{D.~E.} \bibnamefont{Savage}},
  \bibinfo{author}{\bibfnamefont{M.~G.} \bibnamefont{Lagally}},
  \bibinfo{author}{\bibfnamefont{M.}~\bibnamefont{Friesen}},
  \bibinfo{author}{\bibfnamefont{S.~N.} \bibnamefont{Coppersmith}},
  \bibnamefont{and} \bibinfo{author}{\bibfnamefont{M.~A.}
  \bibnamefont{Eriksson}}, High-fidelity resonant gating of a silicon-based
  quantum dot hybrid qubit, \bibinfo{journal}{Npj Quantum Information}
  \textbf{\bibinfo{volume}{1}}, \bibinfo{pages}{1} (\bibinfo{year}{2015}).

\bibitem[{\citenamefont{Schoelkopf et~al.}(1998)\citenamefont{Schoelkopf,
  Wahlgren, Kozhevnikov, Delsing, and Prober}}]{Schoelkopf1998}
\bibinfo{author}{\bibfnamefont{R.~J.} \bibnamefont{Schoelkopf}},
  \bibinfo{author}{\bibfnamefont{P.}~\bibnamefont{Wahlgren}},
  \bibinfo{author}{\bibfnamefont{A.~A.} \bibnamefont{Kozhevnikov}},
  \bibinfo{author}{\bibfnamefont{P.}~\bibnamefont{Delsing}}, \bibnamefont{and}
  \bibinfo{author}{\bibfnamefont{D.~E.} \bibnamefont{Prober}}, The
  Radio-Frequency Single-Electron Transistor (RF-SET): A Fast and
  Ultrasensitive Electrometer, \bibinfo{journal}{Science}
  \textbf{\bibinfo{volume}{280}}, \bibinfo{pages}{1238} (\bibinfo{year}{1998}).

\bibitem[{\citenamefont{Reilly et~al.}(2007)\citenamefont{Reilly, Marcus,
  Hanson, and Gossard}}]{reilly2007fast}
\bibinfo{author}{\bibfnamefont{D.}~\bibnamefont{Reilly}},
  \bibinfo{author}{\bibfnamefont{C.}~\bibnamefont{Marcus}},
  \bibinfo{author}{\bibfnamefont{M.}~\bibnamefont{Hanson}}, \bibnamefont{and}
  \bibinfo{author}{\bibfnamefont{A.}~\bibnamefont{Gossard}}, Fast single-charge
  sensing with a rf quantum point contact, \bibinfo{journal}{Applied Physics
  Letters} \textbf{\bibinfo{volume}{91}}, \bibinfo{pages}{162101}
  (\bibinfo{year}{2007}).

\bibitem[{\citenamefont{Volk et~al.}(2019)\citenamefont{Volk, Chatterjee,
  Ansaloni, Marcus, and Kuemmeth}}]{Volk2019}
\bibinfo{author}{\bibfnamefont{C.}~\bibnamefont{Volk}},
  \bibinfo{author}{\bibfnamefont{A.}~\bibnamefont{Chatterjee}},
  \bibinfo{author}{\bibfnamefont{F.}~\bibnamefont{Ansaloni}},
  \bibinfo{author}{\bibfnamefont{C.~M.} \bibnamefont{Marcus}},
  \bibnamefont{and} \bibinfo{author}{\bibfnamefont{F.}~\bibnamefont{Kuemmeth}},
  Fast Charge Sensing of Si/SiGe Quantum Dots via a High-Frequency Accumulation
  Gate, \bibinfo{journal}{Nano Lett.} \textbf{\bibinfo{volume}{19}},
  \bibinfo{pages}{5628} (\bibinfo{year}{2019}).

\bibitem[{\citenamefont{Connors et~al.}(2020)\citenamefont{Connors, Nelson, and
  Nichol}}]{Connors2020}
\bibinfo{author}{\bibfnamefont{E.~J.} \bibnamefont{Connors}},
  \bibinfo{author}{\bibfnamefont{J.}~\bibnamefont{Nelson}}, \bibnamefont{and}
  \bibinfo{author}{\bibfnamefont{J.~M.} \bibnamefont{Nichol}}, Rapid
  High-Fidelity Spin-State Readout in $\mathrm{Si}$/$\mathrm{Si}$-$\mathrm{Ge}$
  Quantum Dots via rf Reflectometry, \bibinfo{journal}{Phys. Rev. Applied}
  \textbf{\bibinfo{volume}{13}}, \bibinfo{pages}{024019}
  (\bibinfo{year}{2020}).

\bibitem[{\citenamefont{Noiri et~al.}(2020)\citenamefont{Noiri, Takeda, Yoneda,
  Nakajima, Kodera, and Tarucha}}]{noiri2020radio}
\bibinfo{author}{\bibfnamefont{A.}~\bibnamefont{Noiri}},
  \bibinfo{author}{\bibfnamefont{K.}~\bibnamefont{Takeda}},
  \bibinfo{author}{\bibfnamefont{J.}~\bibnamefont{Yoneda}},
  \bibinfo{author}{\bibfnamefont{T.}~\bibnamefont{Nakajima}},
  \bibinfo{author}{\bibfnamefont{T.}~\bibnamefont{Kodera}}, \bibnamefont{and}
  \bibinfo{author}{\bibfnamefont{S.}~\bibnamefont{Tarucha}},
  Radio-Frequency-Detected Fast Charge Sensing in Undoped Silicon Quantum Dots,
  \bibinfo{journal}{Nano Letters} \textbf{\bibinfo{volume}{20}},
  \bibinfo{pages}{947} (\bibinfo{year}{2020}).

\bibitem[{\citenamefont{Taskinen et~al.}(2008)\citenamefont{Taskinen, Starrett,
  Martin, Micolich, Hamilton, Simmons, Ritchie, and
  Pepper}}]{taskinen2008radio}
\bibinfo{author}{\bibfnamefont{L.}~\bibnamefont{Taskinen}},
  \bibinfo{author}{\bibfnamefont{R.}~\bibnamefont{Starrett}},
  \bibinfo{author}{\bibfnamefont{T.}~\bibnamefont{Martin}},
  \bibinfo{author}{\bibfnamefont{A.}~\bibnamefont{Micolich}},
  \bibinfo{author}{\bibfnamefont{A.}~\bibnamefont{Hamilton}},
  \bibinfo{author}{\bibfnamefont{M.}~\bibnamefont{Simmons}},
  \bibinfo{author}{\bibfnamefont{D.}~\bibnamefont{Ritchie}}, \bibnamefont{and}
  \bibinfo{author}{\bibfnamefont{M.}~\bibnamefont{Pepper}}, Radio-frequency
  reflectometry on large gated two-dimensional systems,
  \bibinfo{journal}{Review of Scientific Instruments}
  \textbf{\bibinfo{volume}{79}}, \bibinfo{pages}{123901}
  (\bibinfo{year}{2008}).

\bibitem[{\citenamefont{Samkharadze et~al.}(2016)\citenamefont{Samkharadze,
  Bruno, Scarlino, Zheng, DiVincenzo, DiCarlo, and
  Vandersypen}}]{Samkharadze2016}
\bibinfo{author}{\bibfnamefont{N.}~\bibnamefont{Samkharadze}},
  \bibinfo{author}{\bibfnamefont{A.}~\bibnamefont{Bruno}},
  \bibinfo{author}{\bibfnamefont{P.}~\bibnamefont{Scarlino}},
  \bibinfo{author}{\bibfnamefont{G.}~\bibnamefont{Zheng}},
  \bibinfo{author}{\bibfnamefont{D.~P.} \bibnamefont{DiVincenzo}},
  \bibinfo{author}{\bibfnamefont{L.}~\bibnamefont{DiCarlo}}, \bibnamefont{and}
  \bibinfo{author}{\bibfnamefont{L.~M.~K.} \bibnamefont{Vandersypen}},
  High-Kinetic-Inductance Superconducting Nanowire Resonators for Circuit QED
  in a Magnetic Field, \bibinfo{journal}{Phys. Rev. Applied}
  \textbf{\bibinfo{volume}{5}}, \bibinfo{pages}{044004} (\bibinfo{year}{2016}).

\bibitem[{\citenamefont{Van~Diepen et~al.}(2018)\citenamefont{Van~Diepen,
  Eendebak, Buijtendorp, Mukhopadhyay, Fujita, Reichl, Wegscheider, and
  Vandersypen}}]{van2018automated}
\bibinfo{author}{\bibfnamefont{C.}~\bibnamefont{Van~Diepen}},
  \bibinfo{author}{\bibfnamefont{P.~T.} \bibnamefont{Eendebak}},
  \bibinfo{author}{\bibfnamefont{B.~T.} \bibnamefont{Buijtendorp}},
  \bibinfo{author}{\bibfnamefont{U.}~\bibnamefont{Mukhopadhyay}},
  \bibinfo{author}{\bibfnamefont{T.}~\bibnamefont{Fujita}},
  \bibinfo{author}{\bibfnamefont{C.}~\bibnamefont{Reichl}},
  \bibinfo{author}{\bibfnamefont{W.}~\bibnamefont{Wegscheider}},
  \bibnamefont{and} \bibinfo{author}{\bibfnamefont{L.~M.}
  \bibnamefont{Vandersypen}}, Automated tuning of inter-dot tunnel coupling in
  double quantum dots, \bibinfo{journal}{Applied Physics Letters}
  \textbf{\bibinfo{volume}{113}}, \bibinfo{pages}{033101}
  (\bibinfo{year}{2018}).

\end{thebibliography}

\end{document}